\documentclass[fleqn,usenatbib]{mnras}
\usepackage{newtxtext,newtxmath}
\usepackage[T1]{fontenc}
\usepackage{ae,aecompl}

\usepackage{amsmath}	% Advanced maths commands

\usepackage{graphicx}
\usepackage{subfigure}
\newcommand\msun{{\,M_\odot}}

\newcommand\zsun{{\rm \,Z_\odot}}
\newcommand\lsun{{\rm \,L_\odot}}
\newcommand{\unit}[1]{\ensuremath{\, \mathrm{#1}}}
\usepackage{aecompl}
\usepackage[T1]{fontenc}
\usepackage{threeparttable}
\usepackage{caption}
\usepackage{times}
\usepackage{ifpdf}

\newcommand{\cMpc}{~\mbox{comoving}~\mbox{Mpc}}

\newcommand{\cmci}{~\mbox{cm}^{-3}}

\title[CEMPs in ultra-faint dwarfs]{The role of faint population III supernovae in forming CEMP stars in ultra-faint dwarf galaxies}

\author[Jeon et al.]{Myoungwon Jeon$^{1}$\thanks{E-mail:
myjeon@khu.ac.kr}, 
Volker Bromm$^{2}$, Gurtina Besla$^{3}$, Jinmi Yoon$^{4,5}$, and Yumi Choi$^{6}$\\
$^{1}$School of Space Research, Kyung Hee University, 1732 Deogyeong-daero, Yongin-si, Gyeonggi-do 17104, Korea\\
$^{2}$Department of Astronomy, University of Texas, TX 78712, USA\\
$^{3}$Department of Astronomy, University of Arizona, 933 North Cherry Avenue, Tucson, AZ 85721, USA\\
$^{4}$Department of Physics, University of Notre Dame, Notre Dame, IN 46556, USA\\
$^{5}$Joint Institute for Nuclear Astrophysics—Center for the Evolution of the Elements (JINA-CEE), USA\\
$^{6}$Space Telescope Science Institute, 3700 San Martin Drive, Baltimore, MD 21218, USA
}

\date{Accepted XXX. Received YYY; in original form ZZZ}

\pubyear{2019}

\begin{document}
\label{firstpage}
\pagerange{\pageref{firstpage}--\pageref{lastpage}}
\maketitle

\begin{abstract}
CEMP-no stars, a subset of carbon enhanced metal poor (CEMP) stars ($\rm [C/Fe]\geq0.7$ and $\rm [Fe/H]\lesssim-1$) have been discovered in ultra-faint dwarf (UFD) galaxies, with $M_{\rm vir}\approx10^8\msun$ and $M_{\ast}\approx10^3-10^4\msun$ at $z=0$, as well as in the halo of the Milky Way (MW). These CEMP-no stars are local fossils that may reflect the properties of the first (Pop~III) and second (Pop~II) generation of stars. However, cosmological simulations have struggled to reproduce the observed level of carbon enhancement of the known CEMP-no stars. Here we present new cosmological hydrodynamic zoom-in simulations of isolated UFDs that achieve a gas mass resolution of $m_{\rm gas}\approx60\msun$. We include enrichment from Pop~III faint supernovae (SNe), with $ E_{\rm SN}=0.6\times10^{51}$~erg, to understand the origin of CEMP-no stars. We confirm that Pop~III and Pop~II stars are mainly responsible for the formation of CEMP and C-normal stars respectively. New to this study, we find that a majority of CEMP-no stars in the observed UFDs and the MW halo can be explained by Pop~III SNe with normal explosion energy ($ E_{\rm SN}=1.2\times10^{51}$~erg) and Pop~II enrichment, but faint SNe might also be needed to produce CEMP-no stars with $\rm [C/Fe]\gtrsim2$, corresponding to the absolute carbon abundance of $\rm A(C)\gtrsim6.0$. Furthermore, we find that while we create CEMP-no stars with high carbon ratio $\rm [C/Fe]\approx3-4$, by adopting faint SNe, it is still challenging to reproduce CEMP-no stars with extreme level of carbon abundance of $\rm A(C)\approx 7.0-7.5$, observed both in the MW halo and UFDs.  %It might imply that we should consider less massive progenitor haloes ($M_{\rm vir}<10^8\msun$, $z=0$), in which gas reservoirs available for metal dilution are also small, giving rise to CEMP-no statrs with high $\rm A(C)>7.0$.

\end{abstract}


%%%%%%%%%%%%%%%%%%%%%%%%%%%%%%%%%%%%%%%%%%%%%%%%%%

%%%%%%%%%%%%%%%%% BODY OF PAPER %%%%%%%%%%%%%%%%%%

\section{Introduction}
 
Over the preceding decade, the discovery of metal-poor stars in dwarf galaxies has revolutionized our understanding of the formation and evolution of low-mass dwarf galaxies, with stellar masses $M_{\ast}\lesssim10^6\msun$ at $z=0$ (a recent review is given in \citealt{Simon2019}; see also \citealp{Tolstoy2009, Willman2010, McConnachie2012, Frebel2015}). Stellar archaeology, in particular, allows us to trace back star formation histories (SFHs) of dwarf galaxies, scrutinizing their detailed stellar abundances (e.g. \citealp{Frebel2012, Webster2014, Webster2015, Ji2015}). For instance, [$\alpha$/Fe], the abundance ratio of alpha elements (oxygen, magnesium, calcium, and silicon) compared to iron (e.g. \citealp{Tolstoy2009,Norris2010, Vargas2013}), constitutes a well-known cosmic clock (\citealp{Tinsley1979}), where the break in the [$\alpha$/Fe] vs. [Fe/H] relation marks the onset of Type~Ia supernovae (SNe), thus providing a measure for the overall duration of star formation. 

The carbon to iron ratio, [C/Fe], is often used as another important tracer for stellar populations and early galactic chemical evolution. Many stars observed in the Milky Way (MW) and its satellite dwarf galaxies exhibit enhancement of this ratio, giving rise to a peculiar stellar population called carbon-enhanced metal-poor (CEMP) stars ($\rm [C/Fe]\geq0.7$ and $\rm [Fe/H]\lesssim-1$, see \citealp{Beers2005,Aoki2007,Norris2013} for details). Given the increasing fraction of CEMP stars with decreasing metallicity in the MW halo and MW satellite galaxies (e.g. \citealp{Aoki2007, Frebel2010, Lee2013, Yong2013, Salvadori2015, Placco2014, Yoon2018}), studying CEMP stars with very low-metallicity ([Fe/H] $\lesssim-2.0$) provides great insights into the nature of the early metal-poor star-forming and galactic environments.  Therefore, ultra-faint dwarfs (UFDs), the most metal-poor galaxies with absolute magnitudes fainter than $M_{V}=-7.7$ $(L\lesssim10^5\lsun$; \citealp{Simon2019}), might provide a compelling laboratory to search for hints of CEMP stars (e.g. \citealp{Bland-Hawthorn2015}). In particular, the chemical signature of CEMP-no stars, a subclass of CEMP stars without enhancing heavy neutron-capture process elements ($\rm [Ba/Fe] \lesssim 0.0$), might be unique to the first generation of stars, the so-called Population~III (Pop~III) (e.g. \citealp{Bromm2011,Bromm2013,hansen2016,Yoon2016}). Finding fossil probes of Pop~III stars in Local Group dwarfs is a promising alternative, given the current absence of direct observations of Pop~III stars, which will likely also not be possible with the upcoming {\it James Webb Space Telescope (JWST)} (e.g. \citealp{Gardner2006,Schauer2020}), unless they exploded as particularly superluminous SNe (e.g. \citealp{Whalen2013, Hartwig2018a}).

We focus mainly on the formation of CEMP stars, particularly CEMP-no, as they might preserve the chemical abundance pattern of the first stars. In general, the stellar carbon abundances both in the observed MW halo and its satellite dwarf galaxies are broadly divided into two categories\footnote{We note that this dividing line is somewhat in-definitive due to continuous distribution of [C/Fe] between these populations, though it was defined based on the empirical data \citep{Aoki2007}.}: C-normal ($\rm [C/Fe]<$\ 0.7) and C-enhanced ($\rm [C/Fe]\gtrsim0.7$). The richness of CEMP stars can be found in their $A$\footnote{$A$(C)$=\rm log \epsilon (C)$ is the absolute carbon abundance}(C)-[Fe/H] distribution. \citet{Yoon2016,Yoon2019} recognized that there are three distinct groups (Group~I-III), based on their morphological $A$(C)-[Fe/H] distribution, each of which has a distinct nucleosynthetic signature, suggesting different origins. Most relevant groups to this work are Group II and Group III CEMP-no stars, though Group I CEMP-no with its unusually high $A$(C) value is also of great interest. The Group II stars in the MW and its satellite galaxies appear to be an extension of C-normal stars, whose carbon was co-produced with iron. In contrast, the Group III stars exhibit no correlation between carbon and iron. 

The origin of CEMP-no stars is still actively debated. Possible scenarios include rapidly-rotating Pop~III stars with masses of $m_{\ast}=40-120\msun$ (e.g. \citealp{Meynet2006}), weakly exploding Pop~III SNe with a progenitor mass of $m_{\ast}=10-40\msun$, accompanied by little mixing and low explosion energy (e.g. \citealp{Iwamoto2005, Heger2010, Ishigaki2014}), or jet-like explosions (e.g. \citealp{Tominaga2007, Ezzeddine2019}). Further scenarios propose the mass transfer from an Asymptotic Giant Branch (AGB) star onto a low-mass, metal-poor companion star in binary systems (e.g. \citealp{Suda2004, Abate2013, Starkenburg2014, Arentsen2019}), the efficiency of dust cooling depending on different dust grains properties (e.g. \citealp{Chiaki2017}), or inhomogeneous metal mixing between metal elements (e.g. \citealp{Hartwig2019}). Perhaps, some combination of these scenarios could explain the origins of the distinct CEMP-no groups. Upcoming extensive stellar surveys can provide a more in-depth understanding of their origins.

We focus on the contribution of Pop~III faint SNe on the formation of CEMP-no stars in our simulations. Faint Pop~III SNe have been extensively discussed in many other studies to explain the observed CEMP stars in the local UFDs, but were not considered in our previous study (\citealp{Jeon2017}). These faint SNe, where the weak explosion enables carbon to be released into the interstellar medium (ISM) while iron falls back onto the stellar core, give rise to high intrinsic [C/Fe] ratios (e.g. \citealp{Umeda2003, Iwamoto2005, Tominaga2014, Ishigaki2014, Cooke2014, Komiya2020}). In this study we consider the impact of faint SNe to see if we can reproduce not only Group II but also Group III CEMP-no stars with the extreme levels of carbon enhancement ($\rm [C/Fe]\sim3-4$), which have never been accomplished. 

Further, we investigate the hypothesis by \citet{Yoon2019} that faint Pop~III SNe are likely responsible for the Group III CEMP-no stars with high $A$(C) $\gtrsim6.5$, while the Group II CEMP-no stars in more massive dwarf spheroidal (dSphs) seem to be associated with normal Pop~II SNe. \citet{Yoon2019} argue that higher values of $A$(C) for the CEMP-no stars in local UFDs can be understood, if their birthplaces were chemically primitive low-mass system, where the effects of stochastic and early generation star formation can be maximized as their SFHs are truncated at early stages. This study thus indicates that the UFD ($M_{\rm vir}\lesssim 10^8\msun$ at $z=0$) descendants of low mass dwarfs that host faint Pop III SNe at early times, are the most likely to host CEMP-no stars

Significant theoretical progress has also been made in studying the formation and evolution of dwarf galaxies with cosmological simulations (e.g. \citealp{Sawala2011, Wetzel2015, GK2016}). In particular, hydrodynamic zoom-in simulations have begun to focus on isolated dwarfs ($M_{\rm vir}=10^8-10^9\msun$ at $z=0$), (e.g. \citealp{Simpson2013, Onorbe2015, Wheeler2015, Jeon2017, Fitts2017, Munshi2019, Wright2019}), and only very recently start considering UFDs around MW mass hosts (e.g., \citealp{Applebaum2020}), representing their baryonic component with high resolution. Recently, this approach has been further pushed to the limit of resolving individual star-forming clouds (e.g. \citealp{Rey2019, Rey2020, Wheeler2019, Agertz2020}). 

In our previous work (\citealp{Jeon2017}), we focused on galaxies with $M_{\rm vir}\approx10^9\msun$ at $z=0$, considering typical SN feedback and cosmic reionization, both of which are crucial in suppressing star formation at this mass scale. In addition, we included the initial metal enrichment from Pop~III stars, and compared the simulated dwarfs with local observations. However, these systems were on the borderline between UFDs and classical dSph galaxies, with some systems retaining neutral hydrogen gas to late time (\citealp{Jeon2019}), and exhibiting multiple star formation episodes, analogous to dwarf galaxies such as Leo I and Leo P. Therefore, these are more appropriately analogs to the lowest mass dSphs and dwarf irregular(dIrr) galaxies known.

Here, we build on this work, reaching lower masses to explore the stellar populations and physics that govern the evolution of the lowest mass galaxies known ($M_{\rm vir}\approx10^8\msun$ at $z=0$), with the goal of explaining the origin of the most extreme CEMP-no stars. Achieving high resolution is necessary to resolve the ISM, including dense star forming sites, in such small galaxies. Therefore, we perform simulations with a gas mass resolution of $m_{\rm gas}\sim60\msun$, an order of magnitude enhancement over that in \citet{Jeon2017}. Lastly, compared to \citet{Jeon2017}, we improve the treatment of metal transport, enabling us to distinguish the contributions from Pop~III and Pop~II SNe, by separately tracking individual enrichment from the two populations.

The paper is organized as follows. We describe the numerical methodology in Section~2, and present the detailed simulation results in Section~3, with a focus on the chemical enrichment history of present-day UFD galaxies. We conclude in Section~4 with a summary and some broader perspectives. For consistency, all distances are given in physical (proper) units unless noted otherwise.

\section{Numerical methodology}
\label{Sec:Metho}

\subsection{Simulation Setup}
In this work, simulations have been conducted in a zoomed area, containing six $ M_{\rm vir}\sim10^8\msun$ subhaloes (masses evaluated at $z=0$). To explore how stellar abundance ratios vary depending on which Pop~III SNe explode, we have run two sets of simulations, one using normal SN yields and one including a fraction of faint SNe. In the latter case, we assume that 70$\%$ of Pop~III stars within the $m_{\ast}=10-100\msun$ mass range explode as faint SNe, with the remainder dying as normal SNe. The resulting key characteristics of the simulated galaxies are summarized in Table~1. Note that we here ignore the effects of photoionization heating by stars, which could reduce the fraction of CEMP stars, as star formation in the simulated UFDs can be reduced. We will discuss the impact of photoionization heating on the SFHs and chemical abundances of UFD analogs in a follow-up paper.

\subsection{Gravity, hydrodynamics, and cooling}
We have used a modified version of the N$-$body Smoothed Particle Hydrodynamics (SPH) code {\sc GADGET} (\citealp{Springel2001}; \citealp{Springel2005}), 
to carry out a suite of hydrodynamic zoom-in simulations. We generate the initial conditions using the cosmological initial conditions code {\sc MUSIC} (\citealp{Hahn2011}). We have generated zoom-in initial conditions that include at least six $M_{\rm vir}\sim10^8\msun$ haloes at $z=0$. At first, we conduct a dark matter (DM) only simulation with $128^3$ particles in a $L=3.125 h^{-1} \cMpc$ box to identify the targeted region. Then, four consecutive refinements are carried out, giving rise to an effective resolution of $2048^3$. The dark matter and gas masses in the most refined region are $m_{\rm DM} \approx 500\msun$ and $m_{\rm SPH} \approx 63\msun$, respectively, an order of magnitude improvement over those in \citet{Jeon2017}.
\par

We adopt the following cosmological parameters: a matter density of $\Omega_{\rm m}=1-\Omega_{\Lambda}=0.265$, 
baryon density of $\Omega_{\rm b}=0.0448$, present-day Hubble expansion rate of $H_0 = 71\unit{km\, s^{-1} Mpc^{-1}}$, 
a spectral index $n_{\rm s}=0.963$, and a normalization $\sigma_8=0.8$ (\citealp{Komatsu2011,planck2016}). We use a fixed softening length for DM particles of $\epsilon_{\rm DM}=$ 10 pc
over the entire simulated time, and a variable softening length for gas particles that is proportional to the SPH kernel length with a 
minimum value of $\epsilon_{\rm gas, min}=2.8$ pc. We trace the abundances of nine atomic and molecular species 
($\rm H, H^{+}, H^{-}, H_{2}, H^{+}_2 , He, He^{+}, He^{++},$ and $\rm e^{-}$), as well as the three deuterium species $\rm D, D^{+}$, and HD, under the influence of a 
cosmic UV background \citep{Haardt2011} by solving the coupled, non-equilibrium rate equations every time-step. We gradually introduce the UV background, beginning at 
$z\approx7$, and linearly increase its strength until $z=6$, when cosmic reionization is expected to be complete (e.g. \citealp{Gunn1965}; \citealp{Fan2007}).
\par

The UV background is attenuated through $\exp{(-N_{\rm H~I} \bar{\sigma}_{\rm ion}})$, where the column density of neutral hydrogen is $N_{\rm H~I}=h n_{\rm H~I}$. Here, $h$ is the SPH kernel size and $\bar{\sigma}_{\rm ion}$ the frequency-averaged photoionization cross-section for H~I. Additionally, we account for the photodissociation of molecular hydrogen, $\rm H_2$, by soft UV radiation in the Lyman-Werner (LW) band (11.2 eV$-$13.6 eV) (e.g. \citealp{Abel1997}), with a rate $k_{\rm LW}=1.38\times 10^{-12} {\rm s^{-1}} J_{21}$. Here, the
normalized LW mean intensity is $J_{\rm 21}=\bar{J}_{\nu}/(10^{-21} {\rm erg\, Hz^{-1} s^{-1} cm^{-2} sr^{-1}})$, with $\bar{J}_{\nu}(z)$ being 
the average intensity in the LW band, calculated from the spectra in \citet{Haardt2011}. For simplicity, the treatment for the photodissociation of deuterated hydrogen, HD, is identical to that of $\rm H_2$. 
\par
We take into account all relevant primordial 
cooling processes, including H and He collisional ionization, excitation and recombination cooling, inverse Compton cooling, bremsstrahlung, and collisional excitation cooling of $\rm H_2$ and HD. Also, metal cooling from carbon, oxygen, silicon, magnesium, nitrogen, neon, and iron is considered in their equilibrium states. Specifically, we adopt the method suggested in \citet{Wiersma2009}, where the cooling rates by each species are interpolated as a function of density, temperature, and metallicity from the tables that are computed in advance from the photo-ionization code {\sc CLOUDY} (\citealp{Ferland1998}). Given the small dust content predicted for low-mass galaxies at $z>7$ (\citealt{Jaacks2017}), we here ignore $\rm H_2$ formation through dust and dust cooling.

\begin{table*}
\caption{Characteristics of the simulated UFD analogs at $z=0$ from the simulated region. Two comparison simulations have been performed by adopting two different SN metal yields, corresponding to faint and normal events, respectively. Column (1): Halo name. Column (2): Virial mass (in units of $10^8\msun$).
Column (3): Virial radius (in kpc). Column (4): Stellar mass (in $10^3\msun$). Column (5): Enclosed mass within half stellar mass radius (in $10^6\msun$). 
Column (6): Half stellar mass radius (in pc).
Column (7): Average stellar iron-to-hydrogen ratios. Column (8): Average alpha-to-iron ratios of stars. Column (9): 1-dimensional stellar velocity
dispersion (in km $\rm s^{-1}$). Column (10): Adopted yield table for Pop~III SNe. Note that in some haloes the estimates of $\sigma$ are not available due to the shortage of number of stellar particles within the haloes.}

\centering
\begin{tabular}{c| c| c| c| c| c| c| c| c| c}
\hline
Halo & $M_{\rm vir} $ & $r_{\rm vir} $ & $M_{\ast} $ & $M_{\rm 1/2}$  & $r_{1/2}^{\ast}$ & [Fe/H]& [$\rm \alpha/ Fe$] & $\sigma_{\ast}$ & yield \\ &  $ [10^8\msun]$ & $[\rm kpc]$ & $[10^3\msun]$ & $[10^6\msun]$  &  $[\rm pc]$ & -  & - &  $ [\rm km\, s^{-1}] $ & - \\
\hline
halo1 N & 1.50 & 10.75 & 10 & 1.59 &  181 & -3.05 & 0.81 & 2.04 &  normal SN \\
\hline
halo2 N & 1.36 & 10.45 & 6.67 & 4.35  & 260 & -2.78 & 0.63 &  3.65 &  normal SN \\
\hline
halo3 N & 1.22 & 10.0 & 8.57 & 1.47 & 208 &  -2.75 & 0.86 & 2.75 &  normal SN \\
\hline 
halo4 N & 1.15 & 9.77 & 2.38 & 0.34 &  89 &  -2.22 &  0.59 & N/A  & normal  SN \\
\hline
halo5 N & 1.11 & 9.02 & 9.52& 0.65 & 106 & -2.93  & 0.78 &  3.17  & normal SN \\
\hline
halo6 N & 0.89 & 7.95 & 5.71 & 2.3 & 457 & -2.55 & 0.77 & 2.35 & normal SN \\
\hline

\hline
halo7 F &  1.50 &  10.75 &  3.81& 19.4 & 815 &  -3.29 & 1.47 & 4.15 & faint SN \\
\hline
halo8 F &  1.36 &  10.45 &  5.24 & 1.83 & 160 & -2.84 &  0.90 &  2.41 & faint SN \\
\hline
halo9 F &  1.22 &  10.0 &  7.62 & 0.47 & 91 & -3.06 &  0.90 & 3.59 & faint SN \\
\hline
halo10 F &  1.15 &  9.77 &  0.95& 0.039 & 36 &  -1.90 &  1.18 &  N/A & faint SN \\
\hline
halo11 F &  1.11 &  9.02 &  5.71& 0.11 & 51 & -2.31 &  1.15 & 3.30 &  faint SN \\
\hline
halo12 F &  0.88 &  7.97 &  9.53 & 4.8 & 373 & -3.44 &  1.85 & 3.98 &  faint SN \\
\hline
\end{tabular}
\label{table:simul}
\end{table*}

\subsection{Star formation physics}
Stars are formed from gas clouds at a rate $\dot{\rho}_{\ast}=\rho_{\rm th}/\tau_{\ast}$, where $\tau_{\ast}=\tau_{\rm ff}/\epsilon_{\rm ff}$ is the star formation time scale, $\tau_{\rm ff}=[3\pi/(32G\rho_{\rm th})]^{1/2}$ the free fall time at the threshold density $\rho_{\rm th}$, and $\epsilon_{\rm ff}$ the star formation efficiency per free fall time (\citealp{Schmidt1959}). The star formation efficiency for Pop~III stars has yet to be pinned down precisely (e.g. \citealp{Hirano2015}; \citealp{Stacy2016}). We set $\epsilon_{\rm ff, Pop~III}=\epsilon_{\rm ff, Pop~II}\sim0.01$, which is similar to a typical star formation efficiency in the local Universe. Then, an SPH particle is stochastically converted into a collisionless star particle in a time interval $\Delta t$, if a random number is smaller than min($\Delta t$/$\tau_{\ast}$,1), to follow a given distribution $\dot{\rho}_{\ast}=\rho_{\rm th}/\tau_{\ast}$. The star formation timescale is then given by
\begin{equation}
\tau_{\ast}=\frac{\tau_{\rm ff} (n_{\rm H, th})}{\epsilon_{\rm ff}}\sim400 {\rm Myr} \left(\frac{n_{\rm H, th}}{100 \cmci}\right)^{-1/2}.
\end{equation}

\subsubsection{Population~III}

The characteristic mass of metal-free stars, determined by a complex physical interplay, is still uncertain (e.g. \citealp{Bromm2013}). Compared to the initially predicted masses of typically a few $\sim100\msun$, the mass scale has been shifted to a few $\sim10\msun$ by considering protostellar radiative feedback and disk fragmentation (e.g. \citealp{Stacy2010}), while still extending over a broad range of $1-1000\msun$ (e.g. \citealp{Susa2014}; \citealp{Hirano2015};  \citealp{Hosokawa2016}; \citealp{Stacy2016}). As explained in Section~2.2, the resolution of our simulations ($m_{\rm gas}\approx60\msun$) is high enough to represent individual massive Pop~III stars. We have randomly sampled the mass of individual Pop~III stars from an assumed top-heavy initial mass function (IMF) over a range of $[1-260]\msun$ with a functional form of 
\begin{equation}
\phi = \frac{dN}{d \ln m} \propto m^{-1.3} \exp{\left[ -\left(\frac{m_{\rm char}}{m}\right)^{1.6}\right]}\mbox{\ ,}
\end{equation}
where $m_{\rm char}=30\msun$ is the characteristic mass. Above $m_{\rm char}$, it follows a Salpeter-like IMF, but is exponentially cut off below that mass (e.g. \citealp{Chabrier2003}; \citealp{Wise2012}). We have adopted a density threshold of $n_{\rm th}=100 \cmci$ above which the gas particle is converted into a collisionless star particle by following the star formation prescription described above.
\par

\subsubsection{Population~II}

Second-generation stars, formed out of metal-enriched gas, tend to be long-lived, low-mass stars, unlike the first generation stars. Given their low mass, the resolution here is insufficient to represent individual Pop~II stars. We therefore model Pop~II stars to form as a stellar cluster, comprised of a single stellar population. Two criteria, a density threshold of $n_{\rm th}=100 \cmci$ and a critical metallicity of $Z_{\rm crit}=10^{-5.5}\zsun$, are imposed for the creation of a Pop~II cluster, requiring that gas particles satisfy both conditions. The choice of $Z_{\rm crit}=10^{-5.5}\zsun$ is motivated by dust-continuum cooling (e.g. \citealp{Omukai2000}; \citealp{Bromm2001a}; \citealp{Schneider2010};  \citealp{SSC2016}), where 
dust cooling is responsible for further gas fragmentation at high densities, $n_{\rm H}\gtrsim10^{16}\rm cm^{-3}$, enabling the formation of low-mass stars. Once a gas particle crosses the density threshold, it is converted into a sink particle that immediately accretes surrounding gas until reaching $M_{\ast, \rm Pop~II}=500\msun$, thus representing a Pop~II stellar cluster. For the Pop~II IMF, we adopt a standard Salpeter form, $dN/d\log m\approx m^{-\alpha}$, with a slope $\alpha=1.35$ over the mass range of $[0.1-100]\msun$.

\subsection{Supernova feedback}

In this section, we describe the implementation of the chemical and thermal feedback from SNe. We consider SN explosions from individual Pop~III stars and from Pop~II stellar clusters, separately following the element-by-element metal species from Pop~III and Pop~II nucleosynthesis. In doing so, we can distinguish the level at which a stellar population contributes to the total metal content of a Pop~II cluster. The metals associated with Pop~III stars originate from core-collapse SNe (CCSNe), or powerful pair-instability SNe (PISNe), depending on the initial mass of the progenitor star, which is randomly sampled from the assumed Pop~III IMF. For the Pop~II stellar clusters, metals are released by winds from AGB stars, and/or by Type~II and Type~Ia SNe. The heavy elements from all sources are transported into the neighboring medium via diffusive mixing, enriching the gas clouds out of which subsequent metal-enriched stars are formed. The SN explosion energy is deposited onto the surrounding medium via thermal feedback. We briefly explain the procedure next. 

\subsubsection{Chemical feedback: Population~III stars}

Given that the lack of metals renders mass loss by stellar winds almost negligible (e.g. \citealp{Kudritzki2000}), the final fates of metal-free stars are solely determined by their initial masses (see, e.g.  \citealp{Karlsson2013} for a review). For instance, stellar evolution theory predicts that in the mass range between $10\msun$ and $40\msun$, Pop~III stars are expected to die as CCSNe. Highly energetic PISNe, which are triggered by electron-positron pair creation, are expected for massive stars of $140\msun \lesssim m_{\ast} \lesssim 260\msun$. Very massive stars $m_{\ast}>260\msun$ might undergo direct collapse into a black hole, because the thermonuclear energy release is not sufficient to reverse the infall. We use the values for total ejected masses and metal yields of Pop~III CCSNe from \citet{Heger2010}, who provide nucleosynthetic yields of non-rotating primordial stars in the mass range of $10-100\msun$ by exploring a range of explosion energies of $E_{\rm SN}=[0.3-10]\times10^{52}$ erg and a range of mixing, $\eta=[0.0-0.25]$, between the stellar layers during the explosion.

In our previous work \citep{Jeon2017}, we adopted a metal yield from Pop~III stars with an explosion energy of $E_{\rm SN}=1.2\times10^{51}$ and a mixing efficiency of $\eta=0.015$, which we will refer to as ``normal yield". Provided that faint Pop~III SNe might be mainly responsible for the presence of CEMP stars in the observed local dwarfs (reviewed in \citealp{Nomoto2013}; also see \citealp{Iwamoto2005, Tominaga2007, Ishigaki2014}), we also employ the yield from a faint SN with $E_{\rm SN}=0.6\times10^{51}$ and $\eta=0.06$ (\citealp{Heger2010}). We assume that 70$\%$ of Pop~III stars, with masses eligible for a CCSN death, experience a faint explosion, while the remainder dies as a normal SN, and we refer such mixed occurrence rate to ``faint SN yield". We adopt the yields from \citet{Heger2002} for PISNe. Finally, we only consider faint-SN yields for Pop~III SNe, not for Pop~II SNe.

\subsubsection{Chemical feedback: Population~II stars}

At the end of the life of Pop~II stars, entering the red giant phase, they experience strong mass loss. For instance, intermediate-mass stars ($0.8\msun\lesssim m_{\ast}<8\msun$) might lose up to $60\%$ of their mass during the terminal AGB phase. For AGB metal yields, we employ the values from \citet{Marigo2001}, not considering the kinetic energy input from AGB winds, given that the wind velocity is negligible compared to the velocity dispersion in the ISM. Another metal input from a Pop~II cluster is through Type~II SN explosions of massive stars ($m_{\ast}\gtrsim8\msun$). Here, we utilize the yields from \citet{Portinari1998}, where mass loss on the main sequence is considered. Metals are ejected over multiple timesteps, whenever stars distributed according to the assumed IMF undergo SN explosions within a given hydrodynamic timestep, $\Delta t=0.01-0.1$~Myr. Thus, massive stars inject metals first, followed by relatively less massive stars. For simplicity, the total Type~II SN energy from a single Pop~II cluster is released at once at the end of the most massive star's life in the cluster.
\par
White dwarf remnants are the end-products of low-mass stars with masses below $8\msun$. If such a white dwarf exceeds the Chandrasekhar limit, either by mass transfer from a companion star or the merger of two white dwarfs, a Type~Ia SN is triggered. The large uncertainty originating from the binary evolution makes it difficult to estimate the Type~Ia SN rate, compared to that of Type~II SNe and AGB stars. We adopt empirical delay time functions (e.g. \citealp{Barris2006}; \citealp{Forster2006}), which are characterized by e-folding times, 
$\eta(t)=e^{-t/\tau_{\rm Ia}}/\tau_{\rm Ia}$ where $\tau_{\rm la}=$2 Gyr is the delay time. 
The SN~Ia rate at a given timestep $\Delta t$ is then  $N_{\rm SN~Ia}(t; t+\Delta t) = a \int_t^{t+\Delta t}{f_{\rm wd}(t')\eta(t')dt'}$, where 
$a=0.01$ is a normalization parameter and $f_{\rm wd}$ is the number of stars that have evolved into white dwarfs per unit stellar mass 
(\citealp{Mannucci2006}). We apply the Type~Ia SN spherically symmetric ``W7" model (\citealp{Thielemann2003}) for Type~Ia SN yields. The corresponding SN energy is also deposited onto neighboring particles as thermal energy, but this feedback is likely to be less effective in impacting the surrounding gas because 
the energy injection takes place over billions of years.

\subsubsection{Metal diffusion}

Once metals are ejected from a SN explosion, they are evenly deposited onto neighboring particles, $N_{\rm ngb}=32$, giving rise to the initial metallicity,
$Z_i = m_{\rm metal,i}/(m_{\rm SPH}+m_{\rm metal,i})$,
where $m_{\rm metal,i}=M_{\ast}y_{\rm eff}$. 
Due to the absence of a direct mass flux among SPH particles, an explicit modelling of metal transport is necessary. We utilize a diffusion-based method for the transfer of metals implemented by \citet{Greif2009}, where the diffusion strength is determined by physical quantities at the resolution limit, here the scale of an individual SPH particle (\citealp{Klessen2003}). The diffusion coefficient, defined as $D=2$ $\rho$ $\tilde{v}$ $\tilde{l}$, can be expressed through a characteristic scale, $\tilde{l}$, set by the smoothing length of the SPH kernel, $\tilde{l}=h$, and the density, $\rho$, of the surrounding gas. The velocity dispersion within the kernel,
$\tilde{v}$, is computed via
$\tilde{v}_i^2 = \frac{1}{N_{\rm ngb}} \sum_j |v_i-v_j|^2$, where $v_i$ and $v_j$ are the velocities of particles $i$ and $j$ within the kernel, respectively.

\begin{figure}
    \centering
    \includegraphics[width=8.5cm]{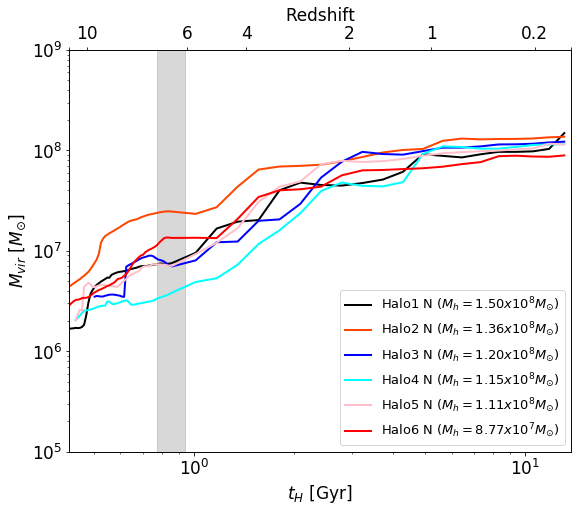}%
    \caption{The growth of the virial mass of the UFDs in the zoomed-in region. The galaxies, initially, start as minihaloes of $M_{\rm vir}\lesssim10^6\msun$ at $z>13$ and evolve into systems of a few $10^8\msun$ at $z=0$, where we find six $M_{\rm vir}\sim10^8\msun$ haloes. We mark the epoch of reionization with the vertical shaded region, where the UV background is introduced at $z=7$, and its strength gradually increases up to $z\approx6$.}
\end{figure}

\begin{figure*}
    \centering
    \includegraphics[width=18cm]{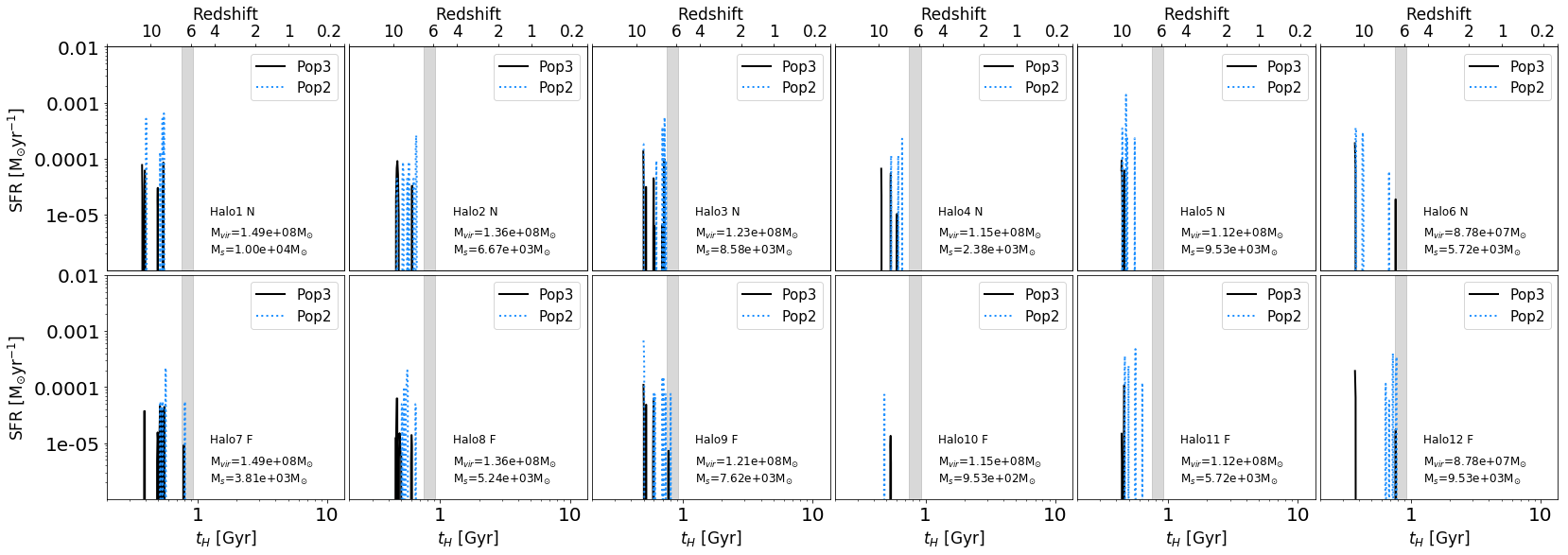}%
    \caption{Star formation rates for Pop~III (black solid lines) and Pop~II (blue dotted lines) as a function of cosmic time in individual galaxies. Note that star formation activities in all progenitor haloes, at a given epoch, are combined. The galaxies in the top panels are from the run with normal SN yields, while 
    the bottom panels exhibit the results for the run assuming faint SN yields. However, any differences between the two runs are due to the stochastic implementation of the star formation process in our model, not the result of using different metal yields. We find that star formation in all simulated UFD analogs is completely quenched both by SN feedback and 
    reionization at $z\sim6$. Some cases, for instance, Halo1~N, Halo5~N, and Halo10~F, indicate self-quenching a few Myr before the onset of reionization, implying that there was not enough time for the gas evacuated by SN feedback to be replenished prior to reionization.}
\end{figure*}

\subsubsection{Thermal feedback}

The SN energy of an individual Pop~III star is released at the end of its lifetime in the form of thermal energy onto the neighboring gas particles. We fix the Pop~III SN energy to $E_{\rm CCSN, Pop~III}=10^{51} \unit{erg}$ and $E_{\rm PISN, Pop~III}=10^{52} \unit{erg}$ for an individual Pop~III CCSN and PISN, respectively. For a Pop~II cluster, the total SN energy, to be injected at once at the death of the cluster's most massive star, is estimated as $E_{\rm CCSN, Pop~II}= \epsilon_{\rm CCSN, Pop~II} \times M_{\ast, \rm Pop~II}=1.8\times10^{52} \unit{erg}$. Here, $\epsilon_{\rm CCSN, Pop~II}=n_{\rm CCSN,Pop~III}\times 10^{51} \unit{erg}$ is the total available CCSN energy per unit stellar mass, and $n_{\rm CCSN,Pop~II}$ the number of CCSNe in a Pop~II cluster for the assumed Salpeter IMF ($[m_0, m_1]=[11,40]\msun$). In general, the number of stars per unit stellar mass resulting in SNe is defined as $n_{\rm SN} = \int_{m_0}^{m_1}{\phi(m)} dm$. Here, $m_0$ and $m_1$ are the minimum and maximum initial mass of stars encountering a SN explosion, and $\phi(m)$ is the given IMF. We use the conventional SN energy of $10^{51}\unit{erg}$ for Pop~II stars. We release the total available SN energy at once, 3~Myr after the formation of a Pop~II cluster, which corresponds to the lifetime of the most massive Pop~II stars.

In order to circumvent the well-known overcooling problem, where the SN energy is rapidly radiated away if the numerical resolution is not sufficient to resolve the SN ejecta evolution (e.g. \citealp{Stinson2007}; Dalla Vecchia \& Schaye 2008; Torrey et al. 2014; Crain et al. 2015; for a comprehensive discussion, see Read et al. 2016a), we employ the method proposed by \citet{Vecchia2012}, in which SN energy is imparted onto only a reduced number of neighboring particles. This ensures a temperature jump of $\Delta T\gtrsim10^{7.5}$K, which is necessary to render SN feedback on the surrounding gas effective. Additionally, we use a timestep-limiter (\citealp{Saitoh2009}, \citealp{Durier2012}) to ensure that the ratio of timesteps of neighboring particles is smaller than a factor of 4, as well as a timestep-update procedure (\citealp{Vecchia2012}), such that neighboring gas particles become active in order to immediately react to a sudden SN energy input.

\section{Simulation results}
In this section, we describe the results obtained from the simulations in terms of two main categories. First, we discuss the SFH of the simulated UFDs and show their resulting global properties. Second, we concentrate on the stellar abundance ratios, and their dependence on the type of Pop~III SNe, normal or faint explosions. In addition, we compare the estimated carbon 
abundances and the CEMP population with observational data.

\subsection{Star formation history}
In the zoomed-in region, there are a total of six $M_{\rm vir}\sim10^8\msun$ galaxies at $z=0$, constituting our sample of UFD analogs. Their progenitor haloes began with $M_{\rm vir} \sim10^6\msun$ DM minihaloes at $z\gtrsim13$, and have grown in mass through mergers and accretion. We have performed two simulations in the zoomed-in region in order to understand how the resulting stellar metal ratios might change, depending on which Pop~III SNe take place. In the first run, all Pop~III stars in the mass range of $10-100 \msun$ die as conventional CCSNe with normal metal yields (Halo1-N$-$Halo6-N). The other run is conducted by assuming that 70$\%$ of Pop~III stars in the CCSN mass range explode as faint SNe, and the remaining 30$\%$ experience normal SN explosions (Halo7-F$-$Halo12-F). Halo properties are listed in Table 1, where haloes are listed in order of decreasing mass. Figure~1 shows the mass growth for Halo1-N$-$Halo6-N, which is identical for Halo7~F$-$Halo12~F.

Given that stars are stochastically formed, the simulated galaxies experience different SFHs. Figure~2 shows the star formation rates (SFRs) as a function of cosmic time for the individual galaxies. We separately track Pop~III and Pop~II star formation, by adding the star formation activities from all the progenitor haloes at a given time. The galaxies in the top panels form in the run with normal SN yields, while the bottom panels illustrate the SFRs arising in the run with faint SN yields. We note that different SFHs should be understood not as a consequence of the SN metal yields used, but reflecting the randomness of the star formation process. SFRs vary from $10^{-6}\msun \rm \,yr^{-1}$ to $10^{-2}\msun \rm \,yr^{-1}$. It can be seen that star formation in all the simulated galaxies is completely quenched both by SN feedback and reionization at $z\sim6$. In some cases, such as Halo1~N, Halo5~N, and Halo10~F, we encounter self-quenching, such that star formation was already inhibited a few Myr prior to the onset of reionization. This is a result of vigorous blow-out of the gas from the host halo by SN feedback, with insufficient time to allow this material to re-collapse.

\begin{figure*}
  \includegraphics[width=130mm]{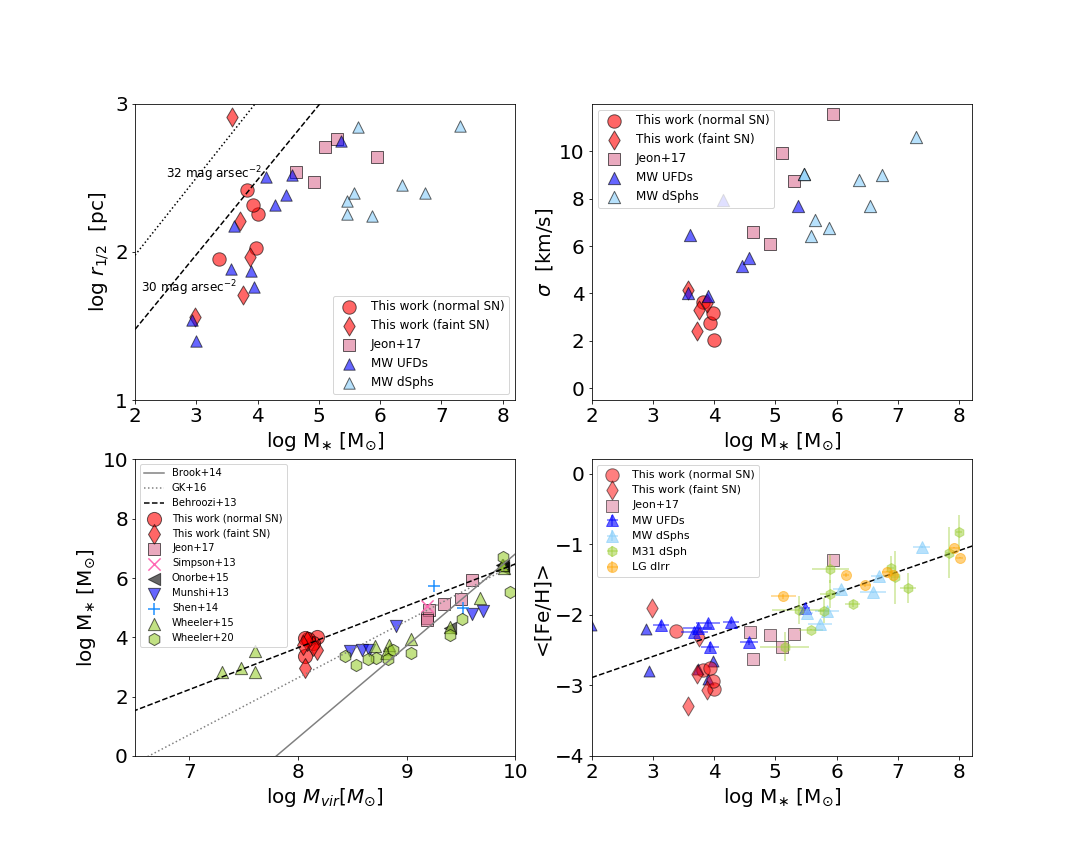}%{fg_evol_pro6.eps}
   \caption{Global properties of the simulated UFD analogs at $z=0$. The panels (clockwise from upper left) show half-mass stellar radius, 1-d velocity dispersion, average stellar metallicity, all as a function of stellar mass, and stellar vs. virial (halo) mass. The resulting quantities for the 12 simulated UFD analogs with normal and faint SN metal yields in this study are shown as red filled circles and diamonds, respectively (see Table 1), and we add the results for both dSph and UFD analogs (pink squares) from \citet{Jeon2017}. The predictions for half-mass stellar radius, velocity dispersion and stellar mass are in good agreement with observed MW UFDs (blue triangles) and MW dSphs (light blue triangles). The stellar vs. halo mass relation for our UFD analogs is in line with the empirical fit provided by \citet{Behroozi2013}. Meanwhile, we here predict average stellar metallicities that are lower by about one dex, compared to the observed UFDs. The reason is that in this work we resolve individual Pop~III stars, enabling Pop~II stars to form immediately after a single Pop~III SN explosion, thus leading to a subset of very low-metallicity Pop~II stars.}
\end{figure*}

%check the number of progenitors!
\begin{figure*}
    \centering
    \includegraphics[width=13cm]{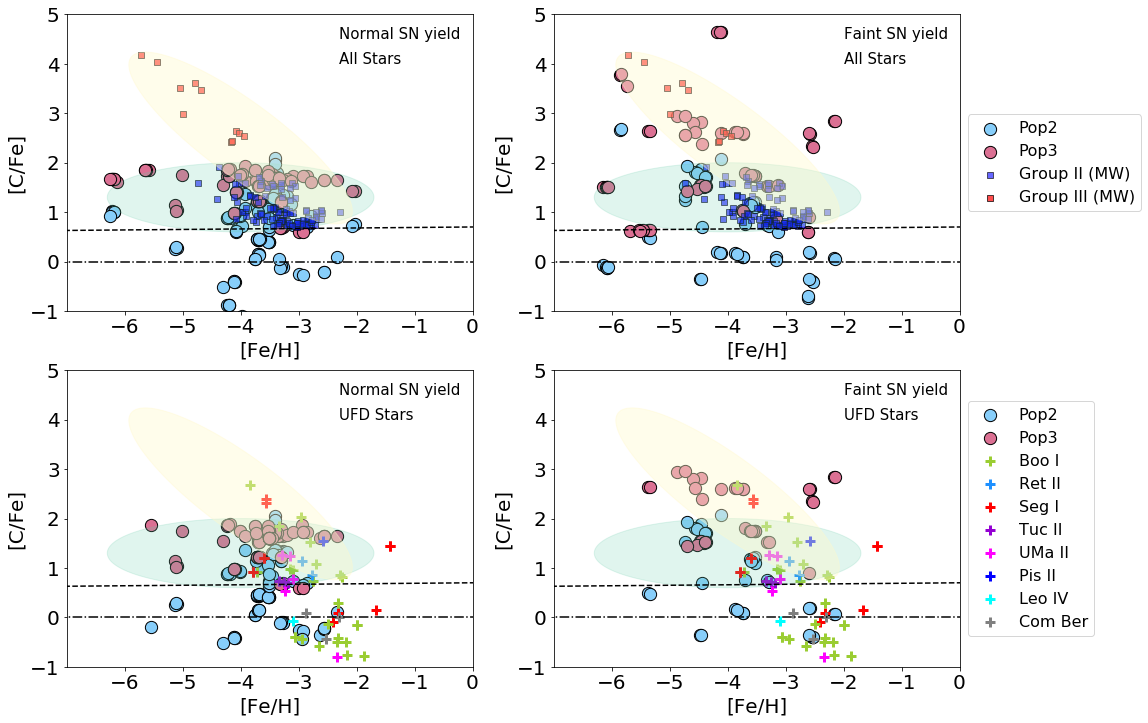}%
    \caption{Stellar carbon-to-iron ratios, [C/Fe], vs. iron abundance, derived from our simulations. {\it Top panels:} All Pop~II stellar clusters, regardless of whether they are associated with UFDs at $z=0$, for the runs with 
    normal (left panel) and faint (right panel) SN metal yields. We distinguish whether metals come from Pop~III (pink circles) or Pop~II (blue circles) SNe, respectively. Evidently, [C/Fe] values from Pop~III SNe tend to be higher than those from Pop~II events, implying that Pop~III stars are mainly responsible for high [C/Fe] ratios. We 
    compare the simulated [C/Fe] values with those observed for CEMP stars in the MW. These stars can be classified into two groups, Group~II (blue squares) and Group~III (red squares), indicated by green and yellow shaded ellipses, respectively (adopted from \citet{Yoon2016}). As can be seen, CEMP stars with $\rm [C/Fe]\gtrsim2.5$, consistent with Group~III, are only produced via the metal yields from faint SNe.
    {\it Bottom panels:} Results for stars residing in the simulated UFD analogs. Here, we compare with [C/Fe] ratios of the observed local UFDs, denoted by colored cross symbols.}   
\end{figure*}

\subsection{Global properties}

We present the global physical properties of the simulated UFD analogs at $z=0$, as summarized in Figure~3. The panels (clockwise from upper left) display the half-mass stellar radius ($r_{1/2}$), 1-dimensional velocity dispersion ($\sigma$), average metallicity (<[Fe/H]>), each as a function of stellar mass, and stellar vs. virial mass. We include quantities both from the two cases with normal (red circles) and faint (red diamonds) SN metal yields. We also exhibit the results from our earlier one in \citet{Jeon2017}. In \citet{Jeon2017} the stellar mass range was $M_{\ast}\approx[4-90]\times10^4\msun$, covering both the observed UFDs and dSphs, whereas here, we only explore the UFD mass range ($M_{\ast}\approx10^3-10^4\msun$). As shown in the top panels, the half-mass stellar radius and velocity dispersion, derived for the simulated UFDs and dSphs, are in line with the values observed for analogous systems in orbit about the MW (\citealp{McConnachie2012}). In the panel for the half-mass stellar radius, we indicate surface brightness limits of 30~mag~arcsec$^{-2}$ and 32~mag~arcsec$^{-2}$. Most of the simulated UFDs are observable with current facilities, whereas some with large half-mass radius are within the fainter limit. We note that the simulated galaxies with faint SNe tend to have a larger range of $r_{1/2}$ than that of the normal SN case. It is understood that this is not because of the difference in the metal yields, but rather due to the randomness of star formation in the systems.

In interpreting the stellar vs. halo (virial) mass relation, we add results from additional zoom-in hydrodynamic simulations (\citealp{Munshi2013, Simpson2013, Shen2014, Onorbe2015, Wheeler2015, Jeon2017, Wheeler2019}), as well as from large-scale simulations, whose results are encapsulated in the best fits, as shown with dashed (\citealp{Behroozi2013}), dotted (\citealp{GK2016}), and solid (\citealp{Brook2014}) lines. The simulated UFDs in this work are well matched to the fit provided by \citet{Behroozi2013}. For the stellar mass$-$metallicity relation, we compare with observational data from \citet{Kirby2013} and \citet{Simon2019}. The average metallicity, derived for our UFD analogs, varies over a broad range, from <[Fe/H]>$\approx-1.9$ to $-3.6$, reaching values that are significantly lower than the observations, and what was found in our earlier work. Our derived range for the total stellar mass in the UFD analogs ($M_{\ast}\approx10^3-10^4\msun$), however, is similar to the observations. 

A possible explanation for the presence of such low metallicity stars, as opposed to our earlier simulations, is the ability to resolve individual Pop~III stars. Here, a single Pop~III SN explosion enables the transition in star formation mode from Pop~III to Pop~II. For example, \citet{Jeon2014} explored Pop~II star formation in detail after such a single Pop~III SN, suggesting that an individual explosion with a progenitor mass of a few $10\msun$ could enrich the ambient gas to a metallicity of $Z=[1-4]\times10^{-4}\zsun$, thus causing the transition to Pop~II star formation. A similar metallicity for initial Pop~II star formation, $\sim 3\times10^{-4}\zsun$, was reported by \citet{Chiaki2019}. The reason for the higher average stellar metallicity in our previous work (\citealp{Jeon2017}; pink squares in Fig.~3) is the numerical representation of Pop~III stars as a stellar cluster, not as individual stars, resulting in the absence of low-metallicity outliers.

Such under-prediction of stellar metallicities in UFD analogs is found by \citet{Wheeler2019} as well. As possible explanations, they suggest the lack of pre-enrichment by Pop~III stars or environmental effects of the more massive host, from which metals could be transferred onto UFDs. The latter is neglected in this work. Alternatively, such low metallicity could be a consequence of unrealistically efficient stellar feedback, which would too strongly disperse enriched gas, resulting in the absence of high metallicity stars. Recently, \citet{Agertz2020} demonstrated that stellar metallicities in UFD-like systems can be reduced by $1-1.5$ dex, when artificially increasing SN energies by 10-100 times the fiducial value. The other global properties, such as $r_{1/2}$, $M_{\ast}$, and $\sigma$, of the simulated UFDs show good agreement with observations. To first order, we can understand this by pointing out that such global quantities are mainly determined by dark matter halo mass, whereas stellar metallicity tends to be sensitive to the adopted subgrid models (\citealp{Agertz2020}).

\subsection{Chemical abundances}
\subsubsection{Carbon}

\paragraph{Pop III vs. Pop II contribution}

Figure~4 shows the resulting [C/Fe] from the runs with normal SNe (left panels) and faint SNe (right panels), comparing with the observational data. We separately present the [C/Fe] ratios of all Pop~II stellar clusters formed in the high-resolution region at $z=0$ (top panels), and the [C/Fe] estimates of Pop~II clusters residing within the virial volume of the UFD analogs (bottom panels). The top panels, therefore, contain more stars, including those in the field, compared to the bottom panels. Furthermore, we consider whether the metals originate from Pop~III (pink circles) or Pop~II (blue circles) SNe, respectively, thus demonstrating that CEMP and C-normal stars are mainly produced through different channels. For instance, it is evident that the [C/Fe] values from Pop~III stars (pink circles) are likely to be higher than those from Pop~II stars (blue circles) by $\sim 1-2$ dex, implying that Pop~III stars are mainly responsible for CEMP signatures with high [C/Fe] ratios. On the other hand, C-normal stars appear to be associated with Pop~II enrichment, shown in blue circles. This trend is more pronounced in the case of faint SNe, where the [C/Fe] ratios from Pop~III SNe can reach up to [C/Fe]$=3-4$, assuming that 70\% of Pop~III stars die as faint events.

When both cases with faint and normal SNe metal yields are combined, we find that mono-enriched Pop~II clusters by Pop~III or Pop~II SNe are 15\% and 5\%, respectively. Here, mono-enrichment is defined as the case where the metal contribution from Pop~III or Pop~II SNe is more than 99\%. Therefore, most of the stars are multi-enriched by metals both from Pop~III and Pop~II SNe, of which 66\% of the Pop~II clusters are predominately polluted by Pop~III SNe meaning that more than 50\% of metals arise from Pop~III SNe. The remaining 34\% of the stars are found to be born out of gas contaminated mainly by Pop~II SNe.

\begin{figure*}
    \centering
    \includegraphics[width=15cm]{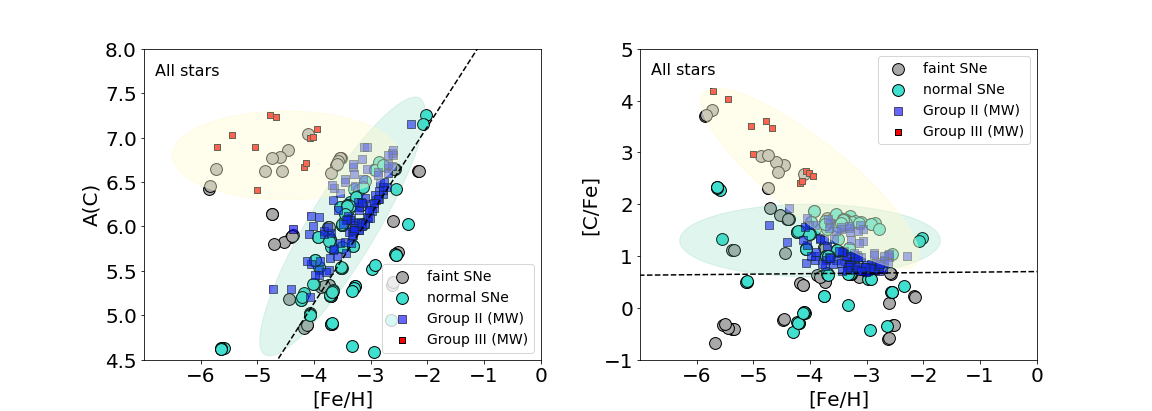}%
    \caption{Carbon abundances as probes of early star formation. $A$(C)-[Fe/H] diagram (left panels) and [C/Fe]-[Fe/H] relation (right panels) of the observed MW stars (red and blue squares) and the stars from the simulations, assuming contributions from normal (green circles) and faint SNe (grey circles). We show all Pop~II clusters found in the zoom-in region, regardless of whether they are bound to the UFDs. The observed CEMP stars in the MW halo can be classified into Group~II (blue squares) and Group~III (red squares), depending on their distinct locations in the $A$(C)-[Fe/H] space. The dashed line corresponds to [C/Fe]=0.7, above which stars are defined as CEMP. Only the stars formed out of gas ejected from faint Pop~III SNe reproduce the stars in the Group~III region (yellow shaded ellipse), as can be seen in both the $A$(C)-[Fe/H] and [C/Fe]-[Fe/H] diagrams.}
\end{figure*}

\begin{figure*}
    \centering
    \includegraphics[width=15cm]{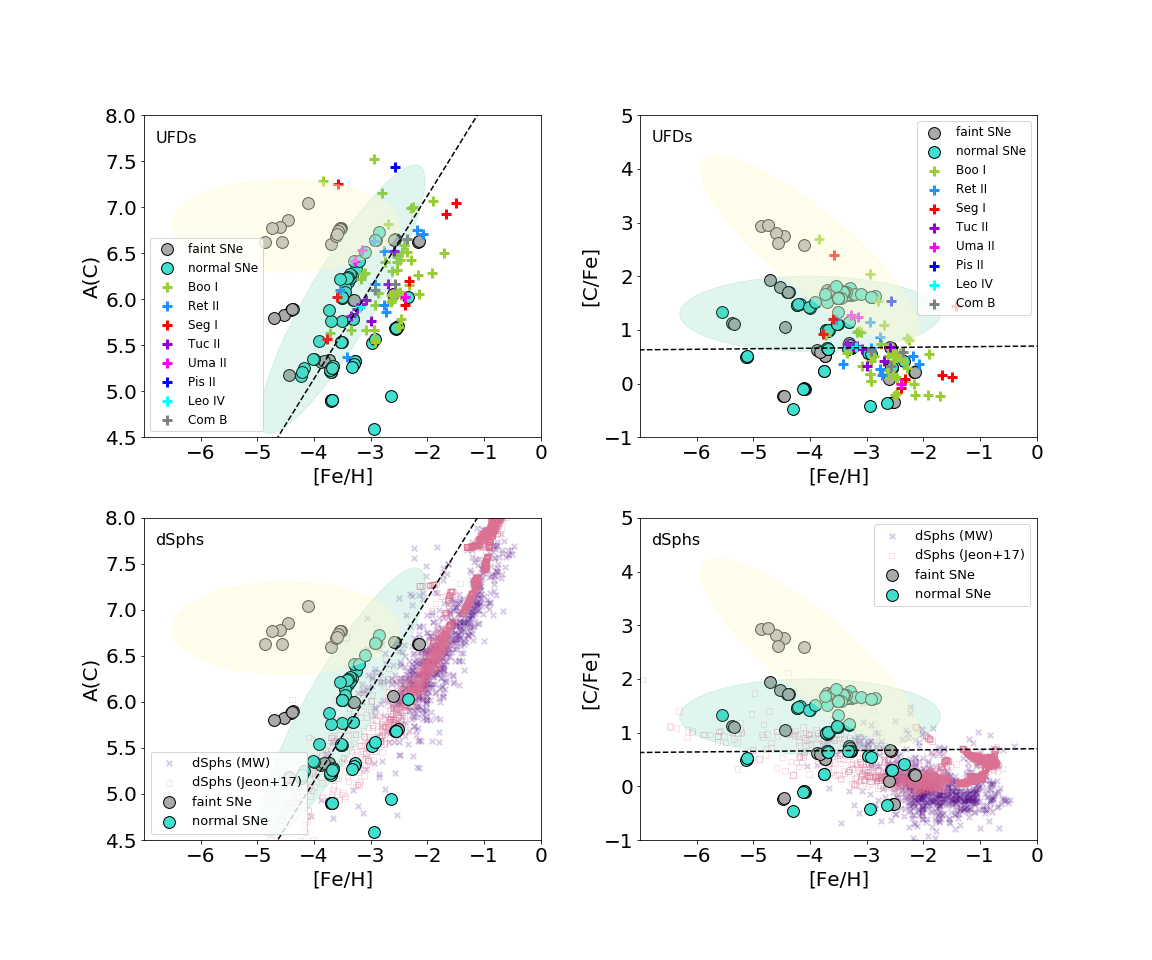}%
    \caption{{\it Top panels:} Same as Fig.~5, but for the subset of stars that lie within the virial volume of the simulated UFDs, comparing with the individual observed UFD stars (crosses with different colors). We note that many Group~III stars in the UFDs lack counterparts in the simulation results, unlike the halo Group~III stars. {\it Bottom panels:} Same as the top panels, but comparing with  the corresponding abundances for the observed dSphs from observations (summarized in \citealt{Yoon2019}, x violet) and simulations (\citealp{Jeon2017}, pink squares). The abundances derived in the present work, together with those of \citet{Jeon2017}, are in overall good agreement with the observational data both for UFDs and dSphs.}
\end{figure*}

\paragraph{Group~II vs. Group~III CEMP stars}

We compare the combined group of all simulated Pop~II clusters with the CEMP stars found in the MW halo. As previously explained, \citet{Yoon2016} have suggested that CEMP stars can be categorized as Group~I (mostly, CEMP-$s$ or CEMP-$r/s\,(i)$ stars, whose carbon is extrinsically enhanced by their AGB binary companion), Group~II ($\rm -5.0 < [Fe/H] < -2.5$ and $5.0 < A(\textrm{C}) < 7.0$), and Group~III ($\rm [Fe/H] < -3.5$ and $ A(\textrm{C}) < 7.5$), based on their distinct location in the $A$(C)-[Fe/H] space. In particular, the latter two groups, mainly comprised of CEMP-no stars, show different behaviors: the absolute carbon abundances of Group~II stars tend to be correlated with [Fe/H], whereas $A$(C) of Group~III stars appears to be uncorrelated with [Fe/H] \citep[see][for more detailed description]{Yoon2016}. \citet{Yoon2019} have suggested, moreover, that the CEMP-no stars in the local UFDs reside in both Groups~II and III regions, while those in more massive dSphs appear to be predominantly Group~II stars.

Figure~4 shows that our simulations can well reproduce CEMP Group~II and Group~III stars, when comparing with those two Group stars from the MW halo CEMP stars and and the MW satellite dwarf galaxies (squares in the top panels) from \citet{Yoon2016}. We note that these observed carbon abundances are evolution corrected values (\citealp{Placco2014}). Interestingly, we find that the stars with the highest ratios of [C/Fe] $>$ 2.5 (right top panel of Figure~4) can only be produced by faint SNe. Also, the trend that [C/Fe] increases with decreasing [Fe/H] appears to be found only in the run with faint SNe, overlapping with the halo Group~III stars (red squares encompassed by the yellow ellipse). Faint SNe are possible for explaining such high [C/Fe] values, due to their weak explosion, where heavy elements are locked up in the core, while light elements such as carbon in the outer layers are released into the ISM. The ambient gas contaminated by this explosion, therefore, is destined to have a large [C/Fe] value, resulting in CEMP star formation. In the bottom panels of Figure~4, we display the [C/Fe] ratios of the stars within the simulated UFD analogs, comparing with those of the observed MW UFDs\footnote{The original references for the abundance data of the UFDs are as follows: \citealp{Norris2010, Lai2011} (Bootes~I); \citealp{Ji2016b, Roederer2016} (Reticulum~II); \citealp{Norris2010, Frebel2014} (Segue~I); \citealp{Ji2016c, Chiti2018a, Chiti2018b} (Tucana~II); \citealp{Frebel2010} (Ursa Minor~II); \citealp{Spite2018} (Pisces~II); \citealp{Simon2010} (Leo~IV); \citealp{Frebel2010} (Coma Berenices).} (crosses in the bottom panels) from \citet{Yoon2019}.

From further inspection of Figure~4, we draw two conclusions. First, it is still valid that Group III CEMP-no stars are associated with Pop~III SNe, while C-normal stars appear to arise preferentially from Pop~II SNe (\citealp{Sarmento2017, Yoon2019}). Both Pop III and Pop II stars are responsible for Group II CEMP stars though CEMP stars with [C/Fe]$\sim 2.0$ appear to be mainly contributed from Pop III SNe. Second, we point out that there are no highly enhanced carbon stars, with [C/Fe]$\approx$3, in the observed UFDs. The reason why we cannot find such stars in the observed UFDs can be understood as follows. The MW halo CEMP-no stars with [C/Fe]$\gtrsim 3$ are all Group~III and rare ultra metal-poor (UMP, [Fe/H]$<-4.0$) stars. Since no UMP stars have been found yet in UFD galaxies, the absence of such extremely carbon enhanced Group~III stars is not surprising. The main observational limitations of finding such stars in the UFDs stem from their faint magnitude and small number statistics per galaxy. The metal-poor halo stars in the MW greatly outnumber the member stars in the UFDs, and thus the probability of finding Group~III-like stars in the UFDs will be correspondingly reduced. Nevertheless, several stars with high [C/Fe] from our simulations are reasonably matched with the values from the run with faint SNe, for instance, one star in Bootes~I ([Fe/H]$\approx-3.84$, [C/Fe]$\approx2.69$), and one in Segue~I ([Fe/H]$\approx$-3.57, [C/Fe]$\approx2.39$).

\paragraph{Faint vs. normal SN yields}

To better understand the nature of CEMP stars in UFDs, we focus on the CEMP Group morphology, exhibited in the $A$(C)-[Fe/H] diagram of \citet{Yoon2016}. In Figure~5 and Figure~6, we compare the resulting $A$(C) absolute abundances (left panels) and [C/Fe] ratios (right panels) from the two runs, adopting faint (grey circles) and normal SN (green circles) metal yields. Figure~5 encompass all Pop~II stars formed in the zoomed region, while Figure~6 presents stars contained only in the simulated UFD analogs. The dashed line corresponds to the [C/Fe]$=0.7$ boundary, above which stars are considered as CEMP. As mentioned earlier, the estimated ratios consistent with Group~III stars, shown as red squares with the averaged value of $A\rm (C)_{avg, obs}\approx6.8$, only arise via faint SN metal yields (grey circles), exhibiting $A\rm (C)_{avg}\approx6.7$. This tendency is also demonstrated in terms of [C/Fe] ratios such that the averaged [C/Fe] ratio of Group~III stars with faint SNe is $\rm [C/Fe]_{avg}\approx3.0$, about 0.3 dex lower than the observed value of $\rm [C/Fe]_{avg, obs}\approx3.3$, whereas it is $\rm [C/Fe]_{avg}\approx2.3$, adopting normal SNe. Both cases with normal and faint SNe show the trend of the CEMP fraction to increase with decreasing metallicity.
About $15-20$\% of the CEMP fraction below $\rm [Fe/H]\approx-2.5$ rises to $\sim 60-70$\% at $\rm [Fe/H]<-5.5$. The behaviour of the CEMP fraction is presented in the Appendix.

Stars in UFDs with high [C/Fe]$\approx2.5-3.0$ ratios (top right panel of Figure~6) can be produced, because relative metal ratios are insensitive to the overall efficiency of the diffusion process. Thus, [C/Fe] ratios are predominantly determined by initial yields of the faint SNe when metals are released, as long as the mixing efficiency for each element is the same. Note that multiple explosions from different progenitor stars, together with the contribution from other metal sources, such as AGB stars and Type~Ia SNe, might modify the resulting [C/Fe]. 

On the other hand, our simulations, even for the faint SN run, fail to reproduce the observed stars in the MW halo and UFDs with high absolute carbon abundance of A(C)$\approx7.0-7.5$. As \citet{Yoon2019} pointed out, UFD systems with smaller gas mass than our current simulations may reproduce such stars because of less dilution of both carbon and iron abundance, leading to higher $A$(C) and [Fe/H] values. Such incapacity of producing CEMP-no stars with high $\rm A(C)>7.0$ might indicate that we should consider UFD analogs even smaller than $\rm M_{\rm vir}<10^8\msun$ ($z=0$), in which gas reservoir eligible for the metal mixing can be minimized. However, we should mention that the mass of gas able to mix with the metals cannot be arbitrarily small. Recently, \citet{Magg2020} demonstrated that the gas enclosed in the final size of the SN remnant could be available for mixing, preventing the metallicity in subsequent stars from becoming too large. Given that the minimum dilution mass suggested by \citet{Magg2020} is based on a single SN explosion, in a case where multiple SN explosions occur in a small halo, we expect that the resulting metallicity can be boosted further despite the fact that there should be the minimum dilution gas mass.

\paragraph{UFDs vs. dSphs}
We also compare the observed $A$(C) and [C/Fe] values from the MW dSphs (data from \citealt{Yoon2019}) with those from the simulated low-mass dSphs by \citet{Jeon2017} in the bottom panels of Figure~6. For the latter, we only plot the stars in the two dSph analogs with $M_{\rm vir}\approx[2-3]\times10^9\msun$ and $M_{\ast}\approx10^5-10^6\msun$ (at $z=0$), values that are 10 (3) times larger than the halo (stellar) masses encountered in this work. Overall, the simulated UFDs in this work and the dSphs in \citet{Jeon2017} show good agreement with the observational data in both the $A$(C)-[Fe/H] and [C/Fe]-[Fe/H] relations. In the $A$(C)-[Fe/H] diagram, the average [Fe/H] of dSphs is shifted toward higher values by $\sim$ 1~dex due to their extended SFHs, as was previously established (e.g. \citealp{Tolstoy2009, Salvadori2015, Yoon2019}). On the other hand, UFDs have experienced relatively short SFHs, at an epoch before the Universe was significantly enriched with metals, thus resulting in the prevalence of metal-poor stars.

\citet{Jeon2017} already suggested that C-normal stars are associated with Pop~II nucleosynthesis, whereas CEMP stars are suggested to arise from normal Pop~III SNe. Note that they only considered normal Pop~III SNe in their work, leading to the absence of CEMP stars with very high ratios of $\rm [C/Fe]>2$. In this work, however, we demonstrate that stars with $\rm [C/Fe]\gtrsim2$ are predominately enriched by metals released by faint SNe. It is still unclear whether the origin of CEMP stars in UFDs and dSphs is fundamentally different (e.g. \citealp{Yoon2019}). However, as UFDs experience truncated SFHs early on, in contrast to those of dSphs, it is highly plausible that stars in UFDs are more likely to retain the signature of Pop~III stars. Furthermore, some of them may have hosted faint SNe, resulting in stars with the unusual signature of very high [C/Fe]. In particular, our simulations confirm the claim in \citet{Yoon2019} that the CEMP stars in dSphs (with $M_{\rm vir}\approx10^9\msun$) predominantly formed from normal CCSN-enriched material, whereas those in UFDs were born out of the ejecta from both normal and faint SNe in their low-mass hosts ($M_{\rm vir}\gtrsim10^6\msun$). This latter formation channel is clearly evident in the bottom panels of Figure~6.

\begin{figure*}
    \centering
    \subfigure{{\includegraphics[width=9.cm]{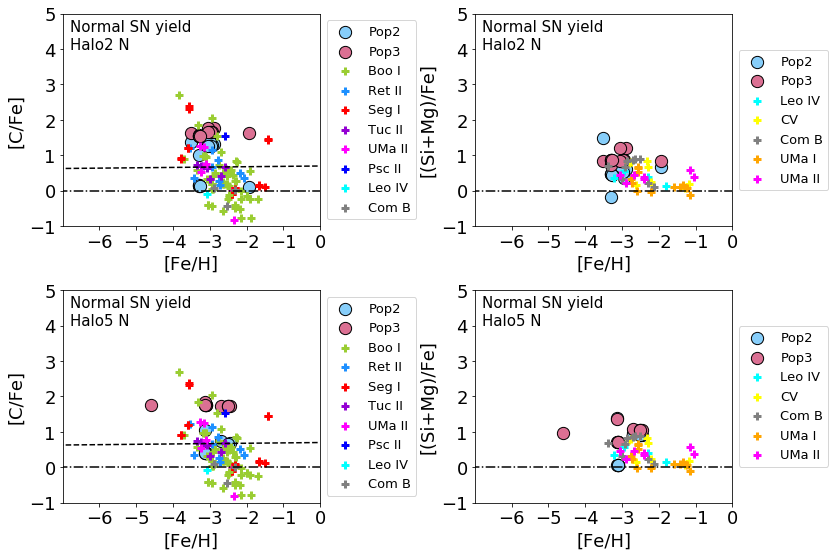} }}%
        \subfigure{{\includegraphics[width=9.cm]{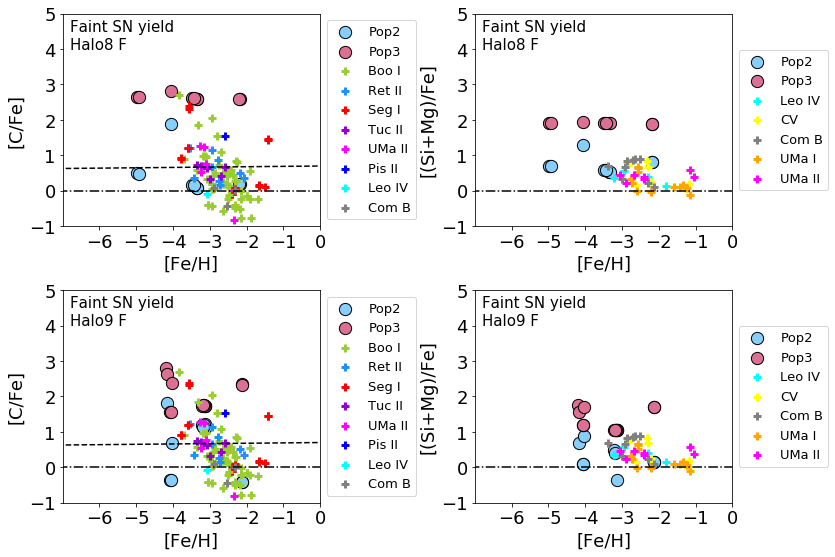} }}%
    \caption{Stellar carbon and $\alpha$-element abundances in select haloes. We consider normal (left two columns) and faint (right two columns) SN metal yields. Compared to the estimates from observed UFDs (cross colored symbols), it appears that metals from Pop~III stars (pink circles) mainly contribute to the composition of CEMP stars. On the other hand, the observed stars with relatively low ratios of [C/Fe] $<0.7$ (C-normal stars) over the metallicity range of $Z\sim10^{-2} - 10^{-3} \zsun$ are in better agreement with the ratios originating from Pop~II stars (blue circles). Meanwhile, assuming faint SN metal yields (right two columns), the derived ratios from the simulations tend to be higher than the ones observed in UFDs by $1-2$~dex.}%
    \label{fig:example}%
\end{figure*}

\paragraph{Comparison with other work}
We note that our results do not imply that our chosen yields uniquely explain the metal abundances of the local UFDs, because we somewhat arbitrarily assign those yields. Although we here distinguish Pop~III SNe as faint or normal simply by their explosion energy, metal yields also depend on additional physics, such as the mixing between stellar layers, even for the same SN energy (e.g. \citealp{Joggerst2009, Heger2010}), or the dimensionality of the assumed nucleosynthesis model (e.g. \citealp{Chen2017}). For our chosen model, the ranges of [C/Fe] ratios that normal and faint SNe metal yields can produce are $\rm [C/Fe]=[-0.43, 2.2]$ and $\rm [C/Fe]=[-0.38, 4.6]$, respectively. As the Pop III progenitor mass increases, the [C/Fe] ratio generally rises. The values, especially, at the characteristic mass of Pop~III stars are $\rm [C/Fe]\approx1.3$ (normal SN) and $\rm [C/Fe]=3.3$ (faint SN). Finding the progenitor masses of the SNe that give rise to CEMP stars requires to match not only the carbon and iron abundance, but also a more comprehensive pattern of heavy elements. As an example, \citet{Placco2016} analyzed metallicity patterns of 20 observed ultra metal poor (UMP) stars in terms of the Pop~III nucleosynthesis model provided by \citet{Heger2010}, suggesting that they can be described by Pop~III stars with masses $m_{\ast}=20.5-28\msun$, exploding as faint SNe ($0.3-0.9\times10^{51}$ erg). 

To the contrary, \citet{Ishigaki2018} argued that these stars are well matched with Pop~III stars of $m_{\ast}\lesssim40\msun$, half of which require hyper-energetic SNe ($\rm E_{SN}\approx10^{52} erg$) with a progenitor mass of $m_{\ast}=25\msun$, based on the fallback-mixing model (e.g. \citealp{Umeda2005, Tominaga2007}). The difference between the two models is that mixing and fallback in the former are achieved within a spherical shell, whereas the latter assumes jet-like fallback, which can lead to strong fallback even for hyper-energetic events. Furthermore, the origin of the two different classes, CEMP and C-normal stars, can be explained in various ways. Here, we account for this dichotomy only in terms of Pop~III vs. Pop~II SN enrichment. Alternatively, the two classes might be the consequence of inhomogeneous mixing of the ejected elements from a single Pop~III explosion with the pristine gas within local ISM (\citealp{Ritter2015, Sluder2016, Hartwig2019}) prior to their formation. 

We note that Group III-like stars have not been yet reproduced by any simulations, even ones considered the inhomogeneous mixing, unlike Group II-like stars (\citealp{Hartwig2019}). Recently, \citealp{Komiya2020} have attempted to reproduce the observed CEMP-no stars using the {\it StarTree}, a merger tree code implementing mixing and fallback models for Pop~III faint SNe. They demonstrated the difficulty of generating the observed CEMP-no stars despite a variety of choices for the model parameters of faint SNe. One of the suggested ways of obtaining CEMP-no stars is to reduce the swept-up mass by SNe and the diffusion coefficient, but it leads to the overproduction of UMP stars compared to the observational data. Such UMP stars could be the possible consequence of Pop~III faint SNe. For instance, \citet{Chiaki2020}, where they explored the degree of pollution by an individual faint SN in a minihalo, showed that a faint SN can enrich nearby gas clouds with $\rm A(C)=3.8-4.9$ at the extremely metal-poor regime, $\rm [Fe/H]\lesssim-8\sim-9$. They explained the predicted $A$(C) values of the gas clouds, lower than the estimates of the observed CEMP-no stars ($\rm A(C)>6$), are attributed to the small metal mass from faint SNe and the fraction of the reincorporated metals into the clouds. However, as they only considered a single Pop~III star formation in a minihalo, multi-enrichment from several SNe events might boost the level of enrichment.

We caution that the observational data set provided by \citet{Yoon2016} should be regarded as an upper limit. According to a recent study by \citet{Norris2019}, when considering the 3-dimensional (3D), local thermodynamic equilibrium (LTE) and 3D, non-LTE (NLTE) corrections for iron and carbon, the estimate of [C/Fe] can be changed significantly. This is because in the 1-dimensional (1D), LTE assumption, carbon is overestimated while iron is underestimated, resulting in a high [C/Fe] ratio. They show that the number of CEMP-no stars ([C/Fe]>0.7, [Fe/H]<-2.0) from \citealp{Yoon2016}, based on 1D LTE, might decrease by 48\% and 73\% due to 3D, LTE and 3D, NLTE effects, respectively. This result, especially, causes a large change in the fraction of Group~II, but the prediction about Group~III stars, which intrinsically have a large amount of carbon, is unlikely to be affected.

\subsubsection{$\alpha$-elements}
In Figure~7, we show two key stellar abundances, [C/Fe] and [$\alpha$/Fe], for the selected individual UFD analogs in the zoomed-in region, assuming normal (left two columns) and faint SN yields (right two columns). Here, the observed UFD $\alpha$ abundances are from \citet{Vargas2013}. Even for individual UFD analogs, the trends found for the combined UFD sample are valid, in that CEMP-no stars mainly arise from Pop~III enrichment, and that [C/Fe] ratios originating from faint SNe are higher by 1-2~dex than from normal SNe. The $\rm [\alpha/Fe]$ ratios also show a similar trend: the average value of <$\rm [\alpha/Fe]$>$=1.24$, when considering faint SNe, is $\sim$0.5~dex higher than the case with normal SN events (<$\rm [\alpha/Fe]$>$=0.74$). 

Compared with our previous work (\citealp{Jeon2017}), adopting the same normal yields, the estimate of <$\rm [\alpha/Fe]$>$=0.74$ for the simulated UFDs in this work is $\sim$0.24~dex higher than the averaged value of <$\rm [\alpha/Fe]$>$=0.50$ for the more massive dwarfs ($M_{\rm vir}\approx10^9\msun$ at $z=0$). This difference implies that the larger the system, the lower the $\rm [\alpha/Fe]$ ratio, as Type~Ia SNe release more iron during their extended SFHs. Finally, we should note that none of the simulated UFD analogs reaches the high metallicity of [Fe/H]$\sim -1$, observed in Ursa Major~I (UMa~I) and Ursa Major~II (UMa~II), although they lie in a similar stellar mass range, $M_{\ast}=1.4\times10^4\msun$ (UMa~I), and $M_{\ast}=4.1\times10^3\msun$ (UMa~II). In the simulations, such metal rich stars, with [Fe/H]$\approx-1$, can be produced when the stellar mass of a galaxy reaches $M_{\ast}\approx8\times10^5\msun$ (\citealp{Jeon2017}). Such metal-rich stars, however, are unlikely to preserve the signature of the initial Pop~II and Pop~III contributions. Instead, their birth clouds were more likely to be contaminated by the nucleosynthetic yields from multiple generations of star formation.

\section{Summary and Conclusions}

We have carried out cosmological hydrodynamic zoom-in simulations to understand the role of faint SNe in the formation of CEMP-no stars in the MW halo and its satellite galaxies. We investigate the simulated galaxies with masses of $M_{\rm vir}\approx10^8\msun (M_{\ast}\lesssim10^{5}$) at $z=0$, which can be considered as direct analogs of UFDs in the local Universe. In order to resolve such low-mass dwarfs, we achieve high resolution with $m_{\rm DM}\simeq 500\msun$ and $m_{\rm gas}\simeq 60\msun$ for dark matter and gas particles, respectively. 

Color-magnitude diagrams of local UFDs indicate that they experienced relatively short SFHs at early times, resulting in only ancient stellar populations (e.g. \citealp{Brown2014, Weisz2014}). As such, UFDs offer a unique window into the characteristics of the first generation of stars, formed a few hundred million years after the Big Bang. CEMP-no stars could be one of the most important empirical features, among metal-poor stars both in the MW halo and UFDs, which might preserve unique signatures of Pop~III stars. In this work, we explore the origin of CEMP-no stars by considering the contribution of Pop~III faint SNe on their birth clouds. We compare specifically the role of faint-SNe with that of the normal-SNe on the formation of CEMP-no stars. By comparing our results with the observed CEMP-no stars in the MW halo and dwarf galaxies, we have shown in this work that such weakly exploding Pop~III SNe provide one of the favored pathways for generating CEMP-no stars. We summarize more specific main findings as follows.

\begin{itemize}

\item The simulated UFDs stop forming stars due to both reionization and SN feedback, leading to short SFHs. Some haloes experience self-quenching solely through SN explosions, halting further star formation already a few Myrs prior to the onset of reionization. Such event occurs if there is not enough time for the evacuated gas due to SN feedback to fall back into its host halo before reionization.
\\
\item The global properties of the simulated UFDs, such as stellar mass, velocity dispersion, and half-mass stellar radius, are in excellent agreement with observational data. However, we predict a broader metallicity range, including values similar to observations, but extending to ones that are lower by about 1.0\, dex.
\\
\item We confirm the results from \citet{Jeon2017} that enrichment from Pop~III stars is mainly responsible for the formation of CEMP-no stars with high ratios of $\rm [C/Fe]\gtrsim2$. In contrast, CEMP stars with $\rm [C/Fe]<2$ including C-normal stars ($\rm [C/Fe]<0.7$) are attributed to the metals ejected from Pop~II stars.
\\
\item The inclusion of faint SN yields enables the formation of stars with very large ratios of $\rm [C/Fe]=2-4$ at $\rm [Fe/H]\lesssim-4$, consistent with the observed stars in the MW halo rather than those in UFDs. Meanwhile, we find that the [C/Fe] vs. [Fe/H] relation for the stars in the observed UFDs can be understood in terms of both faint and normal SNe: [C/Fe], [Fe/H], and $A$(C) values of the stars in the local UFDs are well matched with normal SNe contribution, whereas some stars with high [C/Fe] or $A$(C) require metal yields synthesized via faint SNe. 
\\
\item In the $A$(C)-[Fe/H] diagram, our simulations can reproduce the two CEMP-no sub-groups, Group~II and Group~III (\citealp{Yoon2016}) in the MW halo and satellite dwarf galaxies. Specifically, we reproduce the trend that Group~II stars are predominantly originated from normal SNe and exhibit a correlation between $A$(C) and [Fe/H]. In contrast, the high $A$(C) values ($\gtrsim$ 6.5) of the Group~III stars arise mainly from faint SNe yields and do not have a dependence on [Fe/H]. 
\\
\item Although we find that faint SNe can create Group~III-like stars in terms of $\rm [C/Fe]\approx3-4$ and $\rm A(C)>6.5$ at $\rm [Fe/H]\lesssim-3.5$, it is still difficult to generate CEMP-no stars with extreme level of absolute carbon ($\rm 7.0\lesssim A(C)\lesssim 7.5$), observed both in the MW halo and UFDs. It might imply that we need even less massive haloes ($\rm M_{\rm vir}<10^8\msun, z=0$) to reproduce Group~III-like CEMP-no stars with $\rm A(C)>7.0$. Because less dilution of metals with small gas reservoir would result in CEMP-no stars with high $A$(C) value.
\\
\item The simulation-derived [C/Fe] vs. [Fe/H] and $A$(C) vs. [Fe/H] relations for dSphs from \citet{Jeon2017} are in line with the observed values. The overall shift towards higher metallicity for dSphs is due to their extended SFHs, compared to those of UFDs.
\\

\end{itemize}

Given that detecting the light emitted from the first generation of stars will be extremely challenging even with the upcoming JWST (e.g. \citealp{Magg2018, Liu2020}), stellar archaeology will play a crucial role to study the nature of the first stars, in particular, in local dwarf galaxies. This approach offers an alternative window by providing great details to characterize the first stars and complementary understanding to far-field cosmology. Future observations with greatly improved spectroscopic sensitivity provided by the Giant Magellan Telescope (GMT), the Thirty Meter Telescope (TMT), and the European Extremely Large Telescope (E-ELT) promise to reveal the chemical fingerprint left behind by early cosmic history. Furthermore, metal-poor gas clouds contaminated by the heavy elements released by the first-generation SNe, can be scrutinized for hints to connect the local Universe with that at high-$z$ (e.g. \citealp{Cooke2017, Jeon2019}). This is further complemented by probes of the diffuse intergalactic medium at early cosmic times \citep{Wang2012}. The first stars continue to be elusive but the combined local and high-redshift searches promise to greatly intensify the hunt.

\section*{acknowledgements}
We thank the anonymous referee for constructive and insightful comments that improved the clarity of our paper. We are grateful to Volker Springel, Joop Schaye, and Claudio Dalla
Vecchia for permission to use their versions of \textsc{gadget}. G.~B. acknowledges support from HST Grant 15030. J.Y. acknowledges support by the National Science Foundation under Grant No. PHY-1430152 (JINA Center for the Evolution of the Elements). The simulations were performed using the El Gato cluster at the University of Arizona, which is funded by the National Science Foundation through Grant No. 1228509. We utilized \textsc{yt} for data visualization and analysis tools. M.~J. is supported by the National Research Foundation (NRF) grant No. 20180564, funded by the Korean government (MSIT). The authors acknowledge the Texas Advanced 
Computing Center (TACC) at The University of Texas at Austin for providing 
HPC resources under XSEDE allocation TG-AST120024.
%\footnotesize{}

\section*{DATA AVAILABILITY}
The simulation data and results of this paper may be available upon
request.

\bibliography{myrefs2}{}
\bibliographystyle{mnras} %% mn2e

\appendix
\section{fraction of CEMP stars}
We display in Figure~A1 the fraction of CEMP stars from the two simulations with faint (grey) and normal (blue) SN metal yields, comparing with the observational data (\citealp{Yoon2016}). Note that the observed carbon abundances are evolution corrected values (\citealp{Placco2014}).

\begin{figure}
    \centering
    \includegraphics[width=9cm]{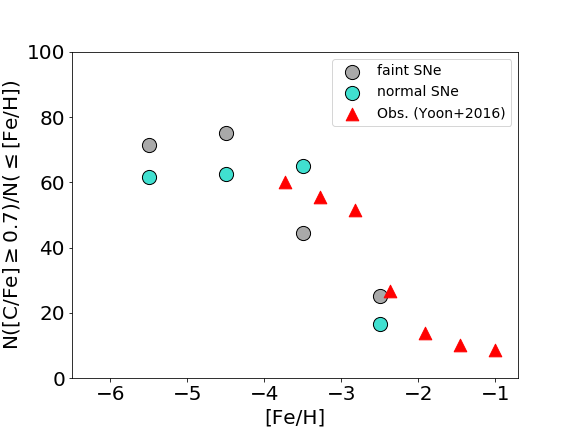}%
    \caption{Cumulative fraction of CEMP stars, defined as $\rm N([C/Fe]\geq0.7) / N (\leq [Fe/H])$, as a function of [Fe/H]. In both simulations with normal and faint SN metal yields, the fraction of CEMP stars tends to increase with decreasing metallicity, showing a good agreement with observational data provided by \citet{Yoon2016}. However, we caution that the observed CEMP fraction can be reduced up to 73\% when 3D, NLTE effects are taken into account (e.g. \citealp{Norris2019}).}
\end{figure}

%%%%%%%%%%%%%%%%%%%%%%%%%%%%%%%%%%%%%%%%%%%%%%%%%%



%\bibliographystyle{mnras}

%%%%%%%%%%%%%%%%%%%%%%%%%%%%%%%%%%%%%%%%%%%%%%%%%%

%%%%%%%%%%%%%%%%% APPENDICES %%%%%%%%%%%%%%%%%%%%%

%%%%%%%%%%%%%%%%%%%%%%%%%%%%%%%%%%%%%%%%%%%%%%%%%%

% Don't change these lines
\bsp	% typesetting comment
\label{lastpage}
\end{document}